\documentclass[twocolumn]{aastex6} 
\usepackage{graphicx}
\usepackage{txfonts}
\usepackage{color}

\fullcollaborationName{The VERITAS Collaboration}

\begin{document} 

\newcommand{\U}{\mbox{4U 0115+634}}
\newcommand{\V}{\mbox{V 404 Cyg}}

   \title{Very-high-energy observations of the binaries \V \ and \U \ during giant X-ray outbursts}
     \shorttitle{VERITAS observation of \V \ and \U}

\author{
A.~Archer\altaffilmark{1},
W.~Benbow\altaffilmark{2},
R.~Bird\altaffilmark{3},
E.~Bourbeau\altaffilmark{4},
M.~Buchovecky\altaffilmark{5},
J.~H.~Buckley\altaffilmark{1},
V.~Bugaev\altaffilmark{1},
K.~Byrum\altaffilmark{6},
M.~Cerruti\altaffilmark{2},
M.~P.~Connolly\altaffilmark{7},
W.~Cui\altaffilmark{8,9},
M.~Errando\altaffilmark{1},
A.~Falcone\altaffilmark{10},
Q.~Feng\altaffilmark{8},
M.~Fernandez-Alonso\altaffilmark{11},
J.~P.~Finley\altaffilmark{8},
H.~Fleischhack\altaffilmark{12},
A.~Flinders\altaffilmark{13},
L.~Fortson\altaffilmark{14},
A.~Furniss\altaffilmark{15},
S.~Griffin\altaffilmark{4},
J.~Grube\altaffilmark{16},
M.~H\"utten\altaffilmark{8},
D.~Hanna\altaffilmark{4},
O.~Hervet\altaffilmark{17},
J.~Holder\altaffilmark{18},
T.~B.~Humensky\altaffilmark{19},
C.~A.~Johnson\altaffilmark{17},
P.~Kaaret\altaffilmark{20},
P.~Kar\altaffilmark{13},
N.~Kelley-Hoskins\altaffilmark{12},
M.~Kertzman\altaffilmark{21},
D.~Kieda\altaffilmark{13},
M.~Krause\altaffilmark{12},
S.~Kumar\altaffilmark{18},
M.~J.~Lang\altaffilmark{7},
T.~T.Y.~Lin\altaffilmark{4},
G.~Maier\altaffilmark{12},
P.~Moriarty\altaffilmark{7},
R.~Mukherjee\altaffilmark{22},
D.~Nieto\altaffilmark{19},
S.~O'Brien\altaffilmark{3},
R.~A.~Ong\altaffilmark{5},
N.~Park\altaffilmark{23},
M.~Pohl\altaffilmark{24,12},
A.~Popkow\altaffilmark{5},
E.~Pueschel\altaffilmark{3},
J.~Quinn\altaffilmark{3},
K.~Ragan\altaffilmark{4},
P.~T.~Reynolds\altaffilmark{25},
G.~T.~Richards\altaffilmark{26},
E.~Roache\altaffilmark{2},
J.~Rousselle\altaffilmark{5},
A.~C.~Rovero\altaffilmark{11},
I.~Sadeh\altaffilmark{12},
S.~Schlenstedt\altaffilmark{12},
G.~H.~Sembroski\altaffilmark{8},
K.~Shahinyan\altaffilmark{14},
D.~Staszak\altaffilmark{23},
I.~Telezhinsky\altaffilmark{24,12},
J.~Tyler\altaffilmark{4},
S.~P.~Wakely\altaffilmark{23},
P.~Wilcox\altaffilmark{20},
A.~Wilhelm\altaffilmark{24,12},
D.~A.~Williams\altaffilmark{17}
}

\altaffiltext{1}{Department of Physics, Washington University, St. Louis, MO 63130, USA}
\altaffiltext{2}{Fred Lawrence Whipple Observatory, Harvard-Smithsonian Center for Astrophysics, Amado, AZ 85645, USA}
\altaffiltext{3}{School of Physics, University College Dublin, Belfield, Dublin 4, Ireland}
\altaffiltext{4}{Physics Department, McGill University, Montreal, QC H3A 2T8, Canada}
\altaffiltext{5}{Department of Physics and Astronomy, University of California, Los Angeles, CA 90095, USA}
\altaffiltext{6}{Argonne National Laboratory, 9700 S. Cass Avenue, Argonne, IL 60439, USA}
\altaffiltext{7}{School of Physics, National University of Ireland Galway, University Road, Galway, Ireland}
\altaffiltext{8}{Department of Physics and Astronomy, Purdue University, West Lafayette, IN 47907, USA}
\altaffiltext{9}{Department of Physics and Center for Astrophysics, Tsinghua University, Beijing 100084, China.}
\altaffiltext{10}{Department of Astronomy and Astrophysics, 525 Davey Lab, Pennsylvania State University, University Park, PA 16802, USA}
\altaffiltext{11}{Instituto de Astronomia y Fisica del Espacio, Casilla de Correo 67 - Sucursal 28, (C1428ZAA) Ciudad Aut—noma de Buenos Aires, Argentina}
\altaffiltext{12}{DESY, Platanenallee 6, 15738 Zeuthen, Germany}
\altaffiltext{13}{Department of Physics and Astronomy, University of Utah, Salt Lake City, UT 84112, USA}
\altaffiltext{14}{School of Physics and Astronomy, University of Minnesota, Minneapolis, MN 55455, USA}
\altaffiltext{15}{Department of Physics, California State University - East Bay, Hayward, CA 94542, USA}
\altaffiltext{16}{Astronomy Department, Adler Planetarium and Astronomy Museum, Chicago, IL 60605, USA}
\altaffiltext{17}{Santa Cruz Institute for Particle Physics and Department of Physics, University of California, Santa Cruz, CA 95064, USA}
\altaffiltext{18}{Department of Physics and Astronomy and the Bartol Research Institute, University of Delaware, Newark, DE 19716, USA}
\altaffiltext{19}{Physics Department, Columbia University, New York, NY 10027, USA}
\altaffiltext{20}{Department of Physics and Astronomy, University of Iowa, Van Allen Hall, Iowa City, IA 52242, USA}
\altaffiltext{21}{Department of Physics and Astronomy, DePauw University, Greencastle, IN 46135-0037, USA}
\altaffiltext{22}{Department of Physics and Astronomy, Barnard College, Columbia University, NY 10027, USA}
\altaffiltext{23}{Enrico Fermi Institute, University of Chicago, Chicago, IL 60637, USA}
\altaffiltext{24}{Institute of Physics and Astronomy, University of Potsdam, 14476 Potsdam-Golm, Germany}
\altaffiltext{25}{Department of Physical Sciences, Cork Institute of Technology, Bishopstown, Cork, Ireland}
\altaffiltext{26}{School of Physics and Center for Relativistic Astrophysics, Georgia Institute of Technology, 837 State Street NW, Atlanta, GA 30332-0430}


\begin{abstract}
Transient X-ray binaries produce major outbursts in which the X-ray flux can increase  over the quiescent level by factors as large as $10^7$.
The low-mass X-ray binary \V \ and the high-mass system \U \ underwent such major outbursts in June and October 2015, respectively. 
We present here observations at energies above hundreds of GeV with the VERITAS observatory taken during some of the brightest X-ray activity ever observed from these systems.
No gamma-ray emission has been detected  by VERITAS in 2.5 hours of observations of the microquasar \V \ from 2015, June 20-21. 
The upper flux limits derived from these observations on the gamma-ray flux above 200 GeV of F $< 4.4\times 10^{-12}$ cm$^{-2}$ s$^{-1}$ correspond to a tiny fraction  (about $10^{-6}$) of the Eddington luminosity of the system, in stark contrast to that seen in the X-ray band.
No gamma rays have been detected during observations of \U \ in the period of major X-ray activity in October 2015. 
The  flux upper limit  derived from our observations is F $< 2.1\times 10^{-12}$ cm$^{-2}$ s$^{-1}$ for gamma rays above 300 GeV, setting an upper limit on the ratio of gamma-ray to X-ray luminosity of less than 4\%.
\end{abstract}


   \maketitle
%

\section{Introduction}

Variable very-high-energy gamma-ray emission ($>$100 GeV, VHE) has been observed from several
X-ray binary systems 
(LS I +61 303, LS  5039, HESS J0632+057, 1FGL J1018.65856, PSR 1259-63, and tentatively Cygnus X-1;
see \cite{Dubus:2013} and \cite{Dubus:2015} for recent reviews). 
All are high-mass X-ray binaries with an O or Be star as stellar companions. 
The observed gamma-ray emission mostly exhibits modulation with the orbital period \citep[e.g.][]{Aharonian:2006}, 
but short intensive flares \citep[e.g.][]{Archambault:2016} have also been observed from some of the gamma-ray binaries.
The origin of the high-energy emission and the acceleration mechanism for the underlying population of relativistic particles 
are not well understood.
They depend, among other features, on the nature of the compact object and the massive star, the geometry of the orbit, and, if applicable, the state of the accretion disk. 
Two major scenarios can be found among the wealth of theoretical work \citep[see][for a review]{Dubus:2013}: 
in the microquasar model, charged particles are accelerated in an accretion-driven relativistic jet, similar to the processes observed in active galactic nuclei. 
Alternatively, the high-energy emission may be driven by shock interaction between a rotation-powered pulsar wind and the strong wind of the massive stellar companion.

Binaries can exhibit huge outbursts during which the X-ray fluxes increase by factors of $10$ - $10^7$ compared to the flux observed during the quiescent stage \citep{Remillard:2006}. 
The systems sometimes reach or even exceed their Eddington luminosity and temporarily become the brightest objects in the X-ray sky. 
In general, the outbursts show no preferred orbital phase and might last for several orbital periods.
VHE emission from X-ray binaries during outbursts has been predicted by a few authors, e.g.~\cite{Orellana:2007}.
Therein, a process is proposed which is based on a mechanism first presented by \cite{Cheng:1989}: a hadronic beam is accelerated in the pulsar magnetosphere and impacts on the transient accretion disk. Neutral pions produced in this interaction decay into a possibly observable flux of gamma rays. 

The known VHE-emitting binaries show strong variability in X-rays. 
The correlation of the VHE emission with the X-ray flux is unclear: some binaries exhibiting clear correlations (e.g.~HESS J0632+057 \cite{Aliu:2014a}), while in others the correlation is not observed during all periods (e.g.~in LS I +61 303,  \cite{Anderhub:2009, Acciari:2011b}). 
LS I +61 303 stands out among the VHE binaries, with strong and short X-ray flares observed with flux doubling times of a few seconds \citep{Smith:2009}. 
For low-mass X-ray binaries, no VHE gamma ray emission has been discovered during giant outbursts. 
However, the sheer amount of energy released in the X-ray range during the giant flares is a strong motivation to observe these types of object at gamma-ray energies \citep{Levinson:1996}.
In this paper,  observations above a few hundred GeV with the VERITAS observatory of the two binaries \V \ and \U \ during major outbursts in 2015 are presented.

\V \ (Ginga 2023+338) is a low-mass X-ray binary system (LMXB) comprised of an evolved K-type star, almost filling its Roche lobe, and a black hole \citep{Casares:1992}. 
The orbital period of the system is 6.47 days, and the binary mass function is f(M)=6.07$\pm$0.05, with typical estimates of the black-hole mass around 9 M$_{\odot}$ or larger \citep{Casares:2014}. 
Its distance of 2.39 $\pm$ 0.14 kpc is very precisely known through parallax measurements \citep{Miller-Jones:2009}.
\V \ is among the brightest LMXBs during quiescence. 
The source underwent transient flaring periods in the optical in 1938, 1956 and 1989.   
X-ray emission was first observed during an outburst in 1989 by the Ginga X-ray observatory  \citep{Makino:1989}. 
The intense, correlated, rapid variability across the spectrum during outbursts is believed to be related to the accretion disk or to jet-like ejection events in the system \citep{Munoz:2016}. 
Similarities with the non-thermal emission in the XRBs Cyg X-1 and GRS 1915+105 have been identified by \cite{Marti:2016} and \cite{Roques:2015}.
Variable X-ray emission in \V\ is also observed during quiescence, the mechanism of which is not well understood \citep{Bernardini:2014}.

On June 15, 2015 at 18:32 UT (MJD 57188.772), the start of a new outburst from \V \ was observed by the X-ray observatory {\it Swift} (GCN \#17929). 
In the hard X-ray band, the source reached unprecedented flux levels, up to 50 times that of the Crab Nebula. 
\cite{Segreto:2015}  estimate the 1-500 keV luminosity during the flare peak to be $1.6\times 10^{39}$ erg s$^{-1}$, consistent with the Eddington luminosity of a 12 M$_{\odot}$ black hole system. The Gamma-ray Burst Monitor (GBM) on board the {\it Fermi} satellite triggered 96 times in 10 days, with the flux reaching 3.7 Crab in the 100-300 keV band on June 19 \citep{Jenke:2015}.
The microquasar was not detected by the Large Area Telescope (LAT) on board the {\it Fermi} satellite during the flare period.
\cite{Siegert:2016} report flux upper limits for the period MJD 57199.616 to 57200.261 of $8\times10^{-7}$  cm$^{-2}$ s$^{-1}$ for gamma rays with energies between 100 MeV and 1 GeV, 
and $3\times10^{-9}$ cm$^{-2}$ s$^{-1}$ in the 1-10 GeV interval.

\U \ is a binary system consisting of an X-ray pulsar and the B0.2Ve star V635 Cas \citep{Reig:2015} in an eccentric (e=0.34) orbit of 24.3 days \citep{Rappaport:1978}.
The period of the pulsar is 3.6 s. 
The distance to the system is estimated to be $6.0\pm1.5$ kpc \citep{Reig:2015}. 
\U \ was discovered during an outburst in 1970 by the Uhuru satellite and has since been observed to have major outbursts approximately every 4-5 years.
The strength and frequency of the outburst are thought to be connected to the timescale of formation and dissipation of the disk of the Be star \citep{Martin:2014}.
The outbursts usually last $\approx 1$ month, with luminosities up to $\approx 10^{38}$ erg s$^{-1}$ \citep[see, e.g.,][]{Tsygankov:2007}. 
The accreting neutron star exhibits several cyclotron resonant-scattering features, which lead to an estimate of the magnetic-field strength of B  $> 10^{13}$ G \citep{Santangelo:1999}.
The Whipple collaboration observed this system in quiescence for 124 hours between 1985 and 1988, and reported a 95\% confidence upper limit of $1 \times 10^{-11}$ cm$^{-2}$ s$^{-1}$ above 0.7 TeV \citep{Macomb:1991}.
The Gas Slit Camera (GSC) of the Monitor of All-sky X-ray Image (MAXI) instrument onboard the International Space Station reported on 2015, October 15 the onset of a new major outburst, reaching a peak of 0.5 Crab flux units in the 15-40 keV band about 7 days after the initial alert \citep{Maxi:2015}.


\section{VERITAS Observatory}

The Very Energetic Radiation Imaging Telescope Array System VERITAS\footnote{\url{http://veritas.sao.arizona.edu/}}
 is a ground-based gamma-ray observatory built to observe high-energy photons
in the energy range from 85 GeV to $>$30 TeV. 
The VERITAS observatory is located at the Fred Lawrence Whipple observatory in southern Arizona, USA
(1.3 km above sea level, N31$^{\mathrm{o}}$40', W 110$^{\mathrm{o}}$57').
It consists of four 12-meter diameter imaging atmospheric Cherenkov telescopes which observe
simultaneously the faint Cherenkov light that is emitted when particle showers initiated by the impact
of high-energy gamma rays pass through the atmosphere.

The flux sensitivity of the instrument reaches 1\% of the flux of the Crab Nebula in 25 hours
in the elevation range relevant for this paper (assuming a detection significance of 5 standard deviations).
The sensitivity of VERITAS to short flares of high-energy emission exceeds significantly the sensitivity of satellite experiments at these energies due to an effective area of $>10^4$ m$^{2}$ above 100 GeV ($>10^5$ m$^{2}$ above 1 TeV).
The angular resolution is better than  0.1 deg at 1 TeV \citep{Holder:2011}. 
Observations with VERITAS are only possible during dark nights and moderate moonlight conditions,
which constrains follow up observations of flaring objects like \U \ and \V.
Some of the observations of \U \ were taken during moonlight conditions,
for which the higher background light levels lead to a lower sensitivity
at the energy threshold \citep{Griffin:2015}.
The analysis results for \U \ are therefore given at a slightly higher energy threshold (300 GeV) than the results on \V\ (200 GeV).

\section{Observations and Results}

%
%


\begin{table*}
\caption{VERITAS Observation Log for \U. The nightly  flux limits are calculated assuming a power-law source spectrum with spectral index of $\Gamma=-2.5$ and for energies above 300 GeV.}   
\label{tab:4UObserving}    
\centering                         
\begin{tabular}{c c cc c c}       
\hline\hline                
Date    & MJD  &  Observing  & Elevation & Observing & Flux upper flux limit  \\
            &  (start of        &   Time          &  range      &  Conditions & (99\% confidence) \\
            &   observations)       &    [minutes]  &                          & & [$10^{-12}$ cm$^{-2}$ s$^{-1}$]    \\
\hline                        
 2015 Oct 23 & 57318.28 & 60 & $56^{\mathrm{o}}$-$57^{\mathrm{o}}$ & bright moonlight & 3.4 \\  
 2015 Oct 23 & 57318.38 & 60 & $45^{\mathrm{o}}$-$49^{\mathrm{o}}$ &  dark & 8.7 \\
 2015 Oct 24 & 57319.25 & 35 & $45^{\mathrm{o}}$-$44^{\mathrm{o}}$ &  bright moonlight & 20.7 \\
 2015 Nov 01 & 57327.16 & 60 & $45^{\mathrm{o}}$-$52^{\mathrm{o}}$ &  dark & 3.8 \\
 2015 Nov 02 & 57328.20 & 60 & $56^{\mathrm{o}}$-$57^{\mathrm{o}}$ &  dark & 3.1 \\
 2015 Nov 03 & 57329.24 & 60 & $56^{\mathrm{o}}$-$57^{\mathrm{o}}$ &  dark & 3.1 \\
\hline                               
\end{tabular}
\end{table*}

 \begin{figure}
   \centering
   \includegraphics[width=\hsize]{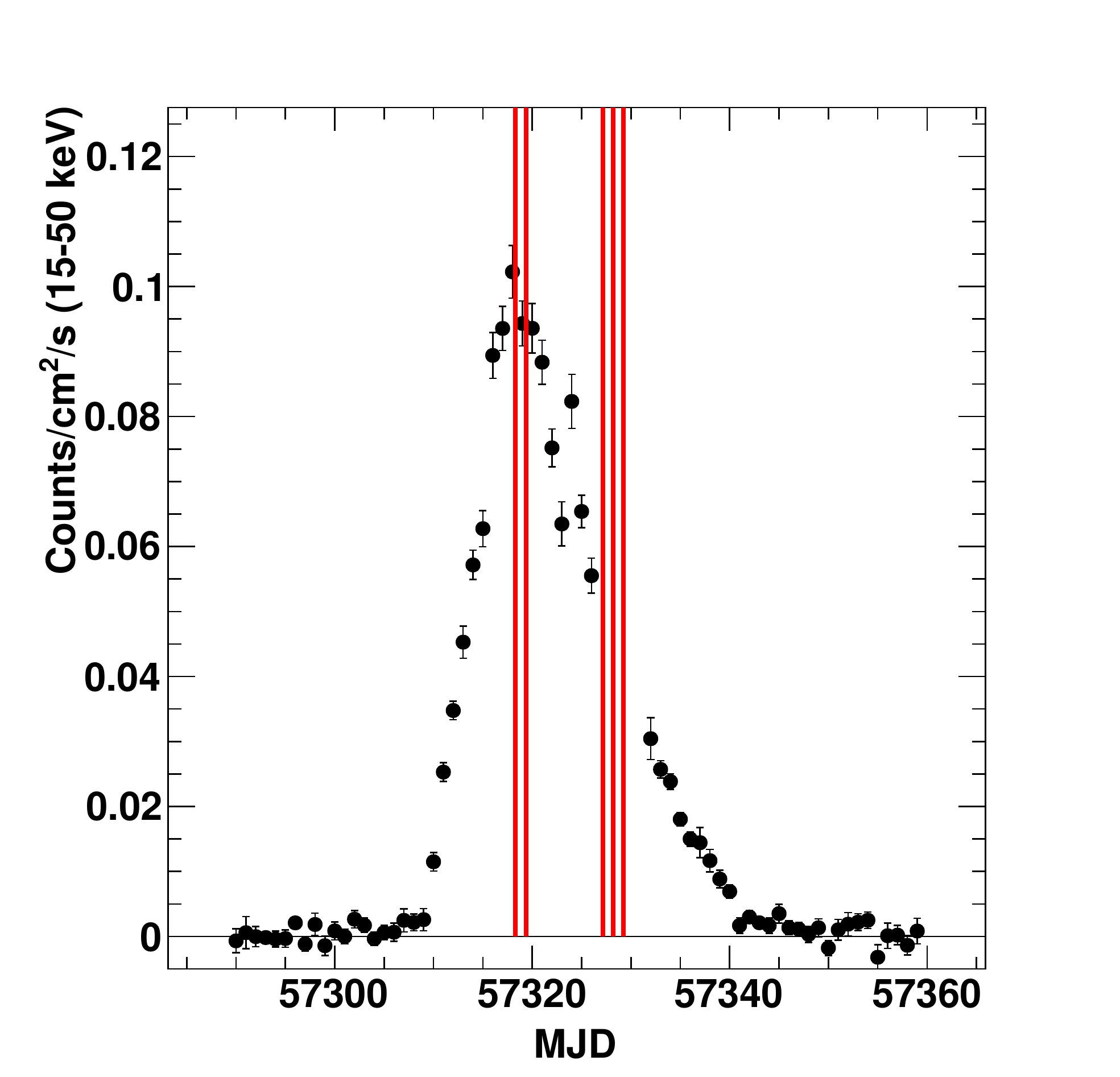}
      \caption{Daily average light curve (15-50 keV) of \U \ for the major outburst in October 2015 obtained with the Burst Alert Telescope (BAT) on board the {\it Swift} satellite. The red lines indicate the time and length of the observing periods with VERITAS.}
         \label{fig:4U0115}
   \end{figure}

%
%


\begin{table*}
\caption{VERITAS Observation Log for \V. The  flux limits are calculated assuming a power-law source spectrum with spectral index of $\Gamma=-2.5$ and for energies above 200 GeV.} 
\label{tab:V4Observing}    
\centering                         
\begin{tabular}{c c c c c c}       
\hline\hline                
Date    & MJD  &  Observing  & Elevation &  Observing & Flux upper limit  \\
            &  (start of        &   Time          &  range      &  Conditions & (99\% confidence) \\
            &    observations)      &    [minutes]  &                  & & [$10^{-12}$ cm$^{-2}$ s$^{-1}$]    \\
\hline                        
 2015 Jun 20 & 57193.38 & 95 & $78^{\mathrm{o}}$-$82^{\mathrm{o}}$ & dark & 5.1 \\  
 2015 Jun 21 & 57194.35 & 60 & $71^{\mathrm{o}}$-$77^{\mathrm{o}}$ & dark & 7.0 \\
 2015 Jun 24 & 57197.34 & 30 & $82^{\mathrm{o}}$-$83^{\mathrm{o}}$ & bad weather, excluded & - \\
\hline                               
\end{tabular}
\end{table*}

 \begin{figure}
   \centering
   \includegraphics[width=\hsize]{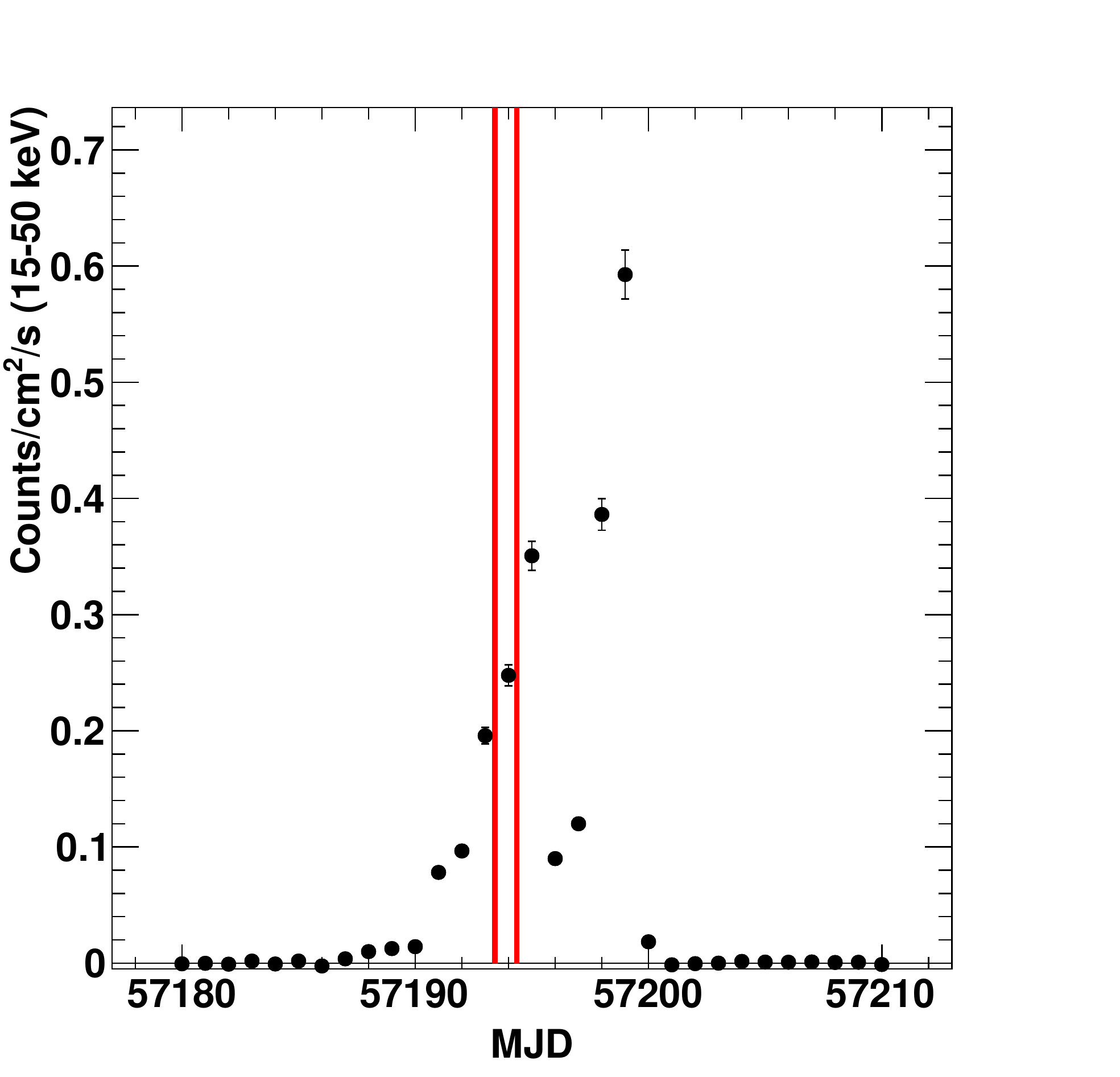}
      \caption{Daily average {\it Swift} BAT light curve (15-50 keV) of \V \ for the major outburst in June 2015. The red lines indicate the time and length of the VERITAS observing periods with good weather conditions.}
         \label{fig:V404Cyg}
   \end{figure}

VERITAS observed \U \ between 2015 Oct 23 and 2015 Nov 3 for about 5.5 hours,
shortly after the first reports of the X-ray outbursts on 2015 Oct 15  \citep{Maxi:2015}.
Figure \ref{fig:4U0115} shows the X-ray light curve from Swift BAT and the VERITAS observing periods.
These are also listed in detail in Table \ref{tab:4UObserving}.
All observing times are given after the application of quality-selection criteria, which remove data
affected by technical problems.
The observations with VERITAS were interrupted by the full-moon period (between 2015 Oct 25 and Oct 31), 
during which the background light levels
are too high for the operation of the sensitive photomultiplier cameras of the observatory.

The search region from the putative gamma-ray source is defined by a circle with radius $0.09^{\mathrm{o}}$.
The background in the search region has been estimated from the same observations using the reflected region model
\citep{Fomin:1994}.
No evidence for VHE gamma-ray emission has been found in the VERITAS data.
The total number of on-source counts $N_{\mathrm{on}}=82$ and off-source counts $N_{\mathrm{off}}=567$
(with a ratio of off-to-on area of $6$) result in a significance of $-1.2\sigma$, following the 
method for significance calculation in \citet[][equation 17]{Li:1983}.
 The flux upper limit at the 99\% confidence level \citep{Rolke:2001} assuming a power-law-like source spectrum with a spectral index 
 of  $\Gamma = -2.5$ is 
 F $< 2.1\times 10^{-12}$ cm$^{-2}$ s$^{-1}$ at energies above 300 GeV
 (corresponding to about 1.6\% of the flux of the Crab Nebula above the same energy).
 Flux upper limits for each observing night can be found in Table \ref{tab:4UObserving}.
 
 
 VERITAS observed \V \ for about 3.3 hours during one of the main outbursts reported at lower energies on the nights 2015 June 20, 21, and 24.
 Figure \ref{fig:V404Cyg} indicates the times of the VERITAS observing periods during the X-ray outburst; all details of the covered time ranges are given in Table \ref{tab:V4Observing}.
 The data taken on 2015 June 24 were taken under inferior weather conditions and therefore excluded from the analysis.
The VERITAS observations were triggered by GCN notification (GCN \#17929) obtained through the rapid-follow up system set in place for observations of gamma-ray bursts \citep{Acciari:2011c}.
The data analysis steps and parameters are exactly the same as for the analysis of \U. 
No evidence for photon emission above an energy  of 200 GeV could be found.
The observed number of events in the signal region is 21, consistent with the expected number of background events of 16.6 derived from 6 background regions located in the same field of view of these observations.
The significance for an excess of gamma rays is $1.0\sigma$.
These numbers correspond to a flux upper limit to the gamma-ray flux above 200 GeV at 99\% confidence level of $F<4.4\times 10^{-12}$ cm$^{-2}$ s$^{-1}$ (corresponding to about 1.9\% of the flux of the Crab Nebula above the same energy).
As \V \ shows  strong variability in X-rays on time scales much shorter than a day \citep{Rodriguez:2015,Kimura:2016}, nightly flux upper limits above the same energy for the two observing nights of VERITAS are listed in Table \ref{tab:V4Observing}.


\section{Conclusions}


%
%
%
The flux upper limits  for \V \  derived from the VERITAS observations and presented in this paper are equivalent to a luminosity of less than $4.0 \times 10^{33}$ erg s$^{-1}$ (above 200 GeV), assuming a distance to \V \ of 2.39 kpc and isotropic emission.
This corresponds to a tiny fraction (about $10^{-6}$) of the Eddington luminosity of the system, in stark contrast with 
e.g.~the  15-200 keV luminosity of $4.6 \times 10^{38}$ erg s$^{-1}$ derived from {\it Swift} XRT measurements during the flare peak \citep{Segreto:2015}.
It is important to note that the X-ray outburst of \V \ is characterised by large flare episodes followed low-flux states, with fluctuations in the X-ray luminosity by more than an order of magnitude.  Luminosities derived from observations by the INTEGRAL Imager on Board the Integral Satellite (IBIS) range from roughly $9 \times 10^{36}$ erg s$^{-1}$ to $4 \times 10^{38}$ erg s$^{-1}$ during the observation periods by VERITAS \citep{Kimura:2016}. 

\V \ shows several similarities to active galactic nuclei (AGN): an accreting black hole and indications of jet-like features.
Most X-rays produced during outbursts in binaries are expected to originate in the accretion disk, while in AGN the synchrotron emission produced by high-energy electrons generally dominates the X-ray flux.
However, hard X-rays (above 10 keV) are generally thought to be an indication for high-energy electrons energised in either a corona or a jet region of binaries.
Evidence for an electron-positron plasma, expected to be present in jets of microquasars, has been observed through the detection of positron annihilation signatures associated with the outburst of \V \  on 2015, June 21 contemporaneously with the VERITAS observations \citep{Siegert:2016}.
The gamma-ray ($>200$ GeV) to hard X-ray (15-50 keV) flux ratio of $<10^{-6}$ derived from the VERITAS measurements of \V \ is orders of magnitude lower than the typical range for these ratios of 0.1-2  observed in gamma-ray bright AGNe \citep[e.g.][]{Aliu:2014, Aleksic:2015}.
 This could mean that no efficient particle acceleration is taking place in these systems during flares, or that essentially all gamma rays are absorbed via pair production with the large number of low-energy photons \citep[see e.g.][]{Dubus:2013, Bednarek:2007}.
The latter process is especially important if particle acceleration takes place close to the base of the jet, i.e.~deeply embedded in the system.
\citet{Orellana:2007} suggested, that X-ray and gamma-ray luminosities are anti-correlated during outbursts, which means that binary systems are most luminous at VHE gamma rays at the beginning or ending of major X-ray outbursts.

%
%

%

\U \ is a pulsar binary system with similarities to the bright gamma-ray binaries LS I +61 303 and HESS J0632+057 (all consisting of Be stars orbited by compact objects on eccentric orbits). 
The flux upper limits derived from the  VERITAS observations for \U \ are equivalent to a luminosity of $<6.4 \times 10^{33}$ erg s$^{-1}$ (above 300 GeV), assuming a distance to \U \ of 6 kpc.
The flux upper limit is not particularly constraining due to the large distance to \U:
it corresponds to about 50\% of the observed gamma-ray luminosity of LS I +61 303 during outbursts near apastron, and is a factor of 3 above the typical luminosity during the gamma-ray bright phases of HESS J0632+057.
%
It is also less constraining then the upper flux limit of $<0.5-1.5 \times 10^{33}$ erg s$^{-1}$ (above 300 GeV) obtained for observations during a giant X-ray flare of the nearby Be/pulsar binary 1 A0535+262 \citep{Acciari:2011}.
However, the observation of \U \ sets a strong upper limit on the ratio of gamma-ray to X-ray luminosity of less than 0.04.
This is significantly lower than the typical ratio of 0.5-1.2 found in the known gamma-ray binaries during their peak emission phase \citep[see Table 2 in][]{Dubus:2013}.
Given these results, and the strong predicted dependency of the gamma-ray emission in Be/pulsar binaries on the properties of the system (e.g. on the inclination angle, size and structure of both the pulsar wind and the stellar wind), \U \ may not be a gamma-ray source comparable to the known gamma-ray binaries.

Future observations with the planned Cherenkov Telescope Array (CTA), with an order of magnitude better  sensitivity than the current gamma-ray instrument, will very likely be necessary to improve further our understanding of the non-thermal processes in these kinds of flaring binary systems.

   
\acknowledgments

   VERITAS is supported by grants from the U.S. Department of Energy Office of Science,
  the U.S. National Science Foundation and the Smithsonian Institution, and by NSERC in
 Canada. We acknowledge the excellent work of the technical support staff at the Fred
 Lawrence Whipple Observatory and at the collaborating institutions in the construction and
 operation of the instrument.
  G.M.~acknowledges support through the Helmholtz Alliance for Astroparticle Physics.
  
 The VERITAS Collaboration is grateful to Trevor Weekes for his seminal contributions
 and leadership in the field of VHE gamma-ray astrophysics, which made this study possible.

\end{document}